\documentclass[runningheads]{llncs}
\usepackage[T1]{fontenc}
\usepackage{graphicx}
\usepackage{tabu} 
\usepackage{url} 
\newcommand\blfootnote[1]{%
  \begingroup
  \renewcommand\thefootnote{}\footnote{#1}%
  \addtocounter{footnote}{-1}%
  \endgroup
}
\begin{document}
\title{Avatar Appearance Beyond Pixels - User Ratings and Avatar Preferences within Health Applications}
\titlerunning{Avatar Appearance Beyond Pixels}
\author{Navid Ashrafi\inst{1,2}\orcidID{0009-0005-8398-415X} \and
Philipp Graf\inst{3}\orcidID{0000-0003-4556-956X} \and
Manuela Marquardt\inst{4}\orcidID{0000-0003-4355-1243} \and
Francesco Vona\inst{1}\orcidID{0000-0003-4558-4989} \and
Julia Schorlemmer\inst{1}\orcidID{0009-0004-7388-9389}\and
Jan-Niklas Voigt-Antons\inst{1}\orcidID{0000-0002-2786-9262} }
\authorrunning{Ashrafi et al.}
\institute{Hamm-Lippstadt University of Applied Sciences\\
\email{name.lastiname@hshl.de} \and
Technical University of Berlin\\
\email{name.lastiname@tu-berlin.de} \and
Munich University of Applied Sciences\\
\email{name.lastiname@hm.edu} \and
Charité University of Berlin\\
\email{name.lastiname@charite.de}}

\maketitle              % typeset the header of the contribution
\blfootnote{This paper has been accepted for publication in the Human-Computer Interaction International Conference 2025. The final authenticated version is available online at \url{https://doi.org/10.1007/978-3-031-93508-4_11}.}
\thispagestyle{empty}
\pagestyle{empty}
\begin{abstract}
The appearance of a virtual avatar significantly influences its perceived appropriateness and the user’s experience, particularly in healthcare applications. This study analyzed interactions with six avatars of varying characteristics in a patient-reported outcome measures (PROMs) application to investigate correlations between avatar ratings and user preferences. Forty-seven participants completed a healthcare survey involving 30 PROMIS items (Global Health and Physical Function) and then rated the avatars on warmth, competence, attractiveness, and human-likeness, as well as their willingness to share personal data. The results showed that competence was the most critical factor in avatar selection, while human-likeness had minimal impact on health data disclosure. Gender did not significantly affect the ratings, but clothing style played a key role, with male avatars in professional attire rated higher in competence due to gender-stereotypical expectations. In contrast, professional female avatars were rated lower in warmth and attractiveness. These findings underline the importance of thoughtful avatar design in healthcare applications to enhance user experience and engagement.

\end{abstract}
\section{Introduction}
Virtual Agents (VAs) or avatars are digital characters simulating a human-like interaction to create a more familiar and appealing experience for the users \cite{Guimaraes2020-xs}. Avatars are designed to fulfill user needs and preferences and can communicate via voice, gestural, and facial expressions. Nowadays, VAs have become increasingly popular in various domains such as healthcare, education, and entertainment \cite{Norouzi2018-uu}. In healthcare, virtual avatars could be utilized to educate and inform patients and caregivers, provide emotional support and social presence to patients, assist in medical consultation and therapy, prescribe medicine, or be used as orthopedic trainers to instruct exercises \cite{Milne-Ives2020-rq}. Therefore, creating an appropriate and well-matched avatar is significant for a healthcare application's overall effectiveness and adoption and directly influences the users' quality of experience (QoE). Moreover, avatars play a crucial role in enhancing the acceptability and satisfaction of eHealth solutions and can emulate human-like interactions, such as interviews \cite{franch}. As a result, there is a growing body of research dedicated to understanding the essential features, optimal design practices, and assessment metrics for evaluating the success rate of healthcare applications utilizing virtual avatars \cite{second}.

However, despite their potential usability and efficiency, using virtual avatars within healthcare settings and clinics is not common practice. This issue is closely linked to the current lack of research on the ethical impact that such systems can have in clinical care. While usability and reliability are widely addressed by researchers, only a limited number of studies explicitly explore the trustworthiness and the acceptability of avatars in healthcare. Other important factors influencing user engagement and happiness with an avatar are how the user perceives the avatar's characteristics, such as warmth, competence, and attractiveness. Although the precise determinants behind selecting a particular avatar remain incompletely understood \cite{Fong}, beyond personal preferences and requirements, the specific usage scenario of the application could directly shape the perceived suitability of a particular avatar's appearance and the selection of VAs. Hence, in this research work, we have analyzed the ratings and choice of avatars with different appearances within a healthcare application containing Patient-Reported Outcome Measures (PROM). PROMs are validated surveys standardized to evaluate patients' perceptions of their health status, encompassing aspects such as pain levels, mobility, and the ability to perform daily routines or specific tasks. PROMs are usually collected at several points in time during a patient's care program. They are beneficial for monitoring a patient's progress and help caregivers have a better understanding of their patient's health status. In this data disclosure scenario, virtual avatars can provide companionship and guidance for the patients while filling out the PROM questionnaires.

This work investigates the implication of utilizing virtual agents in non-pharmacological medical treatments concerning the appearance of the agents. For the purpose of this study, we have developed an application with six various avatars interacting with participants to assist them in filling in a PROM questionnaire containing general and mental health-related items. Each user interacted with all avatars in a randomized order. We then asked the participants to rate their interaction with each avatar and choose the agent they found most desirable and suitable for such PROM applications. Furthermore, we have looked into the factors that influence the specific choice of avatar and trustworthiness of this type of virtual assistant within our PROM application. This research contributes to investigating the perceived avatar appearance within healthcare applications and aims to enhance the QoE provided by such systems.

\begin{figure}[htbp]
\includegraphics[width=8cm, height=9cm]{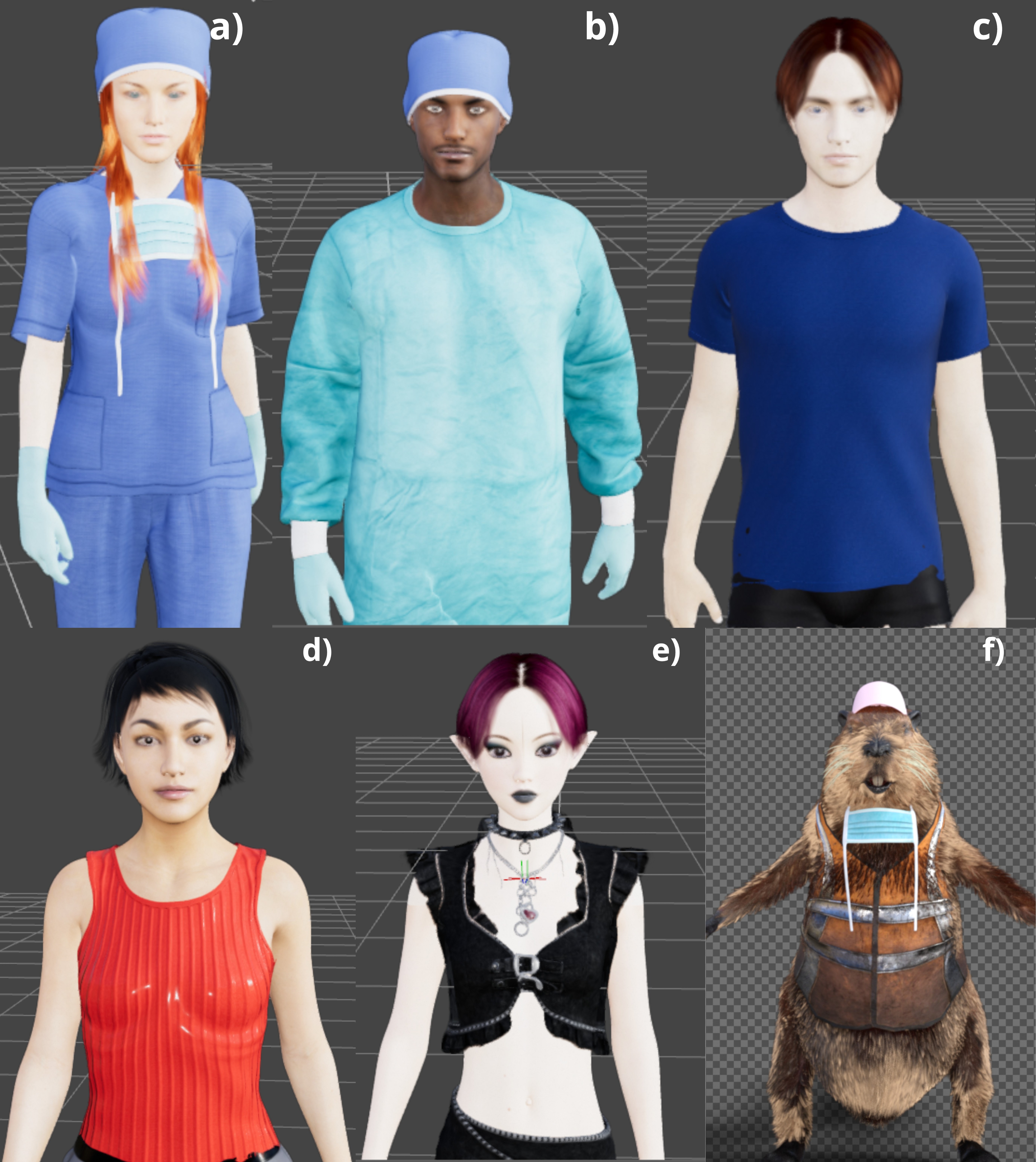}
\centering
%\captionsetup{justification=centering}
\caption{Virtual agents used for the experiment with different gender and social roles. Top left: Avatar \textit{A} [F,P], top middle: Avatar \textit{B} [M,P], top right: Avatar \textit{C} [M,C], buttom left: Avatar \textit{D} [F,C], buttom middle: Avatar \textit{E} [F,O], and buttom right Avatar \textit{F} [O] (M: Male, F: Female, C: Casual, P: Professional, O: Others)}
\label{teaser}
\end{figure}

\section{Related work}

Virtual avatars - when appropriately customized - can potentially motivate and encourage patients to fulfill their routine healthcare tasks, such as completing surveys. Avatars can be used as an efficient alternative for human forces in healthcare with regard to performing repetitive tasks that are excessively time and energy-consuming. They play an important role in enhancing the QoE in connection with availability and accessibility to serve the high demand for information and interaction among patients. Besides classical control inputs, biofeedback can enhance the interaction in immersive media systems \cite{kojic2019impact}. The video recording from faces can be used to estimate the emotional state of system users \cite{porcu2020estimation}.

The collection of health data electronically (Meirte et al. 2020) and assisted by agent-based technologies also harbours potential for reducing barriers and removing infrastructural hindrances (Long et al. 2022). Avatars can also be utilized as an initial point of contact before redirecting the patient to a human or medical professional. Although avatars cannot completely simulate a human-like medical interaction, they imitate patterns and scripts of human interaction when engaging with patients and can potentially utilize the sense of companionship within mental healthcare applications \cite{Emi}. Likewise in mental healthcare, the usage of embodied virtual agents potentially improves the quality of treatment and reduces costs. However, the success rate of incorporating avatars in mental healthcare by medical experts and patients is still not fully discovered \cite{shaikh}. As the extensive literature review by ter Stal et al. \cite{ter2020design} on embodied conversational agents used in eHealth applications shows, there is still disagreement on which design features the virtual agents should include in such scenarios.

One of the main concerns surrounding the interaction with VAs in mental healthcare is the self-disclosure of patients. Namely, the patient's willingness to share personal information, and also thoughts, feelings, and uncomfortable experiences, with the goal of delivering a proper diagnosis \cite{archer,fiske}. Studies on this issue have shown that virtual avatars potentially encourage users to self-disclose by providing verbal and nonverbal cues in their communication \cite{33,32}. %Virtual agents have been recently employed to help users make healthier lifestyle choices and increase their mental health \cite{14,30,50}, as well as motivate patients towards a more honest interaction.

In \cite{41}, authors developed a chatbot that offered conversationally relevant self-disclosures in real-time using a large conversation dataset, engaging users with active social conversations. Another research \cite{moon} investigated the influence of the wording of questions on the responses of the participants and discovered that self-disclosure of the interviewer resulted in participants exchanging more intimate information. Although research on self-disclosure using chatbots and virtual avatars has experienced significant growth \cite{stock2023tell}, the main research questions, such as how virtual avatars can contribute to deep self-disclosure in healthcare, are still unanswered \cite{lee}. 

Furthermore, warmth and competence are two main social factors that shape our perception and evaluation of individuals. Warmth and competence provide insight into a person's goals and their endeavor to pursue them \cite{fiske1}. Although avatars are virtual non-human characters, they are often treated as if they were real social agents \cite{moon,nass}. Therefore, users tend to assess avatars according to their perceived warmth and competence, similar to how they assess real people.
Perceived warmth and competence are also important in rating an avatar's trustworthiness. People tend to prefer avatars that are categorized as more warm and competent since they feel more credibility and trust in these avatars \cite{10}.  Another research from ter Stal et al. \cite{ter2020you} reported that people have a tendency to trust avatars that have the same age range and gender as them and Parmar et al. \cite{parmar2018looking} showed that an avatar's attire influences users' trust. Rheu et al.  \cite{rheu2021systematic} reviewed articles on conversational agents, which revealed five key design themes influencing trust: social intelligence, voice characteristics, communication style, agent appearance, non-verbal communication, and performance quality. Furthermore, the study noted that the demographic factors, personality traits, and context of use play a role in moderating the impact of these design themes on trust. Moreover, different avatar characteristics may resonate within the user's personality, e.g., a friendly avatar might fulfill a user's need for relatedness, as characteristics represented in a particular avatar are often extended to the user who picked it \cite{NOWAk}. Therefore, when investigating the underlying factors contributing to a specific avatar choice, it is important to consider user demographics data and -if possible, emotional and psychological background. 

Although there are definitely additional factors that lead to a particular avatar choice, evidence shows that need satisfaction plays a critical role to the extent that it could derive game choices and behavior, indicating that it probably also influences avatar selection \cite{12,13}. Users may choose avatars that would reflect their individual preferences, including their psychological needs \cite{11}. In this research work, we have designed six avatars with different characteristics and attire to serve as virtual assistants within a healthcare questionnaire application and had 47 users interacting and rating all the avatars. Our primary aim was to answer the question of which avatar suits such applications better and to analyze the factors influencing user ratings. This work contributes to the current body of research regarding virtual avatar design for healthcare and paves the way for further future work in this area.

\section{Methodology}

%Prior to this study we have the following hypotheses:
%\begin{itemize}
%\item {\verb|acmsmall|}: People with lower GMHS would be more likely to pick an avatar that is perceived as more warm and competent by them.
%\item {\verb|acmlarge|}: Users would be more willing to share their personal information rather with the avatar they could feel more comfortable communicating with.
%\end{itemize}
%To test our hypothesis, 
%To investigate the underlying factors contributing to avatar choice, we conducted a study where participants interacted with several different avatars and rated their interaction with the avatars, as well as rated each avatar's characteristics separately.

In this section, we will go through a detailed procedure of our experiment, followed by a brief introduction to our participants and recruitment process. Furthermore, we will provide an overview of the application design, including the sculpting and designing of the avatars and the scenes. Finally, we will present the PROM items and avatar rating questionnaires used in this study.

\subsection{Process of the Experiment}

\begin{figure}[htbp]
\centerline{\includegraphics[width=8cm, height=8cm]{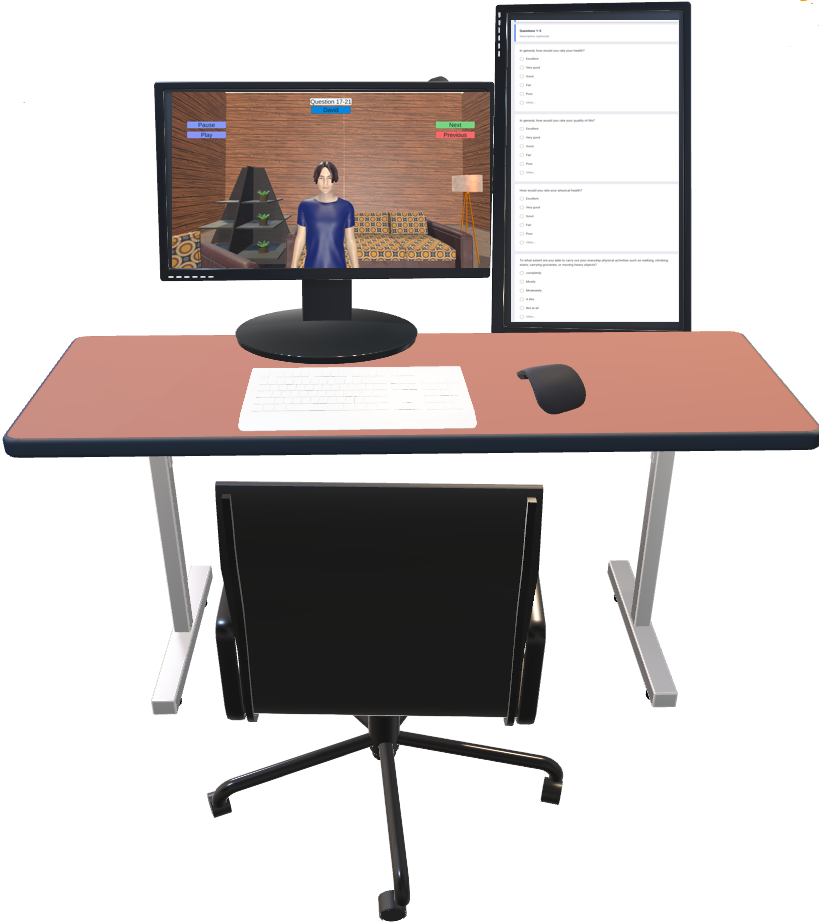}}
\centering
%\captionsetup{justification=centering}
\caption{Experiment setup including two adjacent monitors for displaying the scene with the avatars (on the left desktop) and the questionnaires (on the right desktop).}
\label{setup}
\end{figure}

The setup of the experiment included two external screens (Figure \ref{setup}). One monitor on the left side would display the avatars and the surrounding scene (in 4K resolution) in landscape mode. We designed the application such that six different avatars would appear in a randomized order in the same scene environment. The screen on the right-hand side would display (in portrait mode) several questionnaire tabs, including one health-related questionnaire, one avatar-rating questionnaire per avatar, and one final questionnaire to pick the favorite avatar among the six (Figure \ref{setup}). The health questionnaire consisted of 30 items from the Patient-Reported Outcomes Measurement Information System (PROMIS). Each avatar would then read out loud five questions to the users, such that the users could answer the questions while listening to the avatar simultaneously or after the avatar has read the question. In each scene, the avatar would be positioned standing in the center of the scene looking into the camera. Five seconds of delay between reading each question was set by default to give the user enough time to answer the questions. After every five questions, the user was asked to open the questionnaire tab with the avatar name on the right-side screen to rate their interaction with the avatar, their willingness to share information, perceived warmth, perceived competence, perceived attractiveness, and human-likeness of the avatar. The user would then proceed with the next set of PROM items with the next avatar on the left-side screen. Additionally, an open text box was provided to collect user comments and feedback on each avatar rating, and also on the final avatar-choice questionnaire.

\subsection{Participants}

Participants were recruited through the university study advertising platforms in two different locations (Berlin \& Lippstadt, Germany). In total, 47 participants were included in the analysis, consisting of 28 males, 18 females, and one identifying as non-binary (mean age 27, range 18-45). The users were given an informed consent form to read and sign, confirming their voluntary participation in the study and explaining the data privacy policy and the freedom to skip items in the questionnaire or to quit the study at any time. The participants were then given a short 5-minute description of the study by the examiner. The examiner would then start the experiment program, open the questionnaire tabs for the participants, and let them begin the experiment. The examiner would stay in the room for the duration of the experiment in case the user needed support. The examiner would not monitor the user input while filling in the surveys. The implementation of the experiment was approved by the ethics committee of the University of Applied Sciences Hamm-Lippstadt. 

\subsection{Scene Design}

The avatar application was developed using the \textit{Unity} game engine \cite{unity2021}. We created a cozy-looking hut environment with wooden walls, couches, tables, and plants to immersive the user in a relaxing and comfortable space. The same environment was used for all scenes, with a different camera angle from time to time. To design the avatars, \textit{Daz Studio} \cite{daz2022} was chosen. \textit{Daz Studio} is a 3D modeling tool specialized in providing rigged 3D human models. We used the \textit{Genesis 8.1} base characters - with built-in face morph and lip-synch capabilities - to sculpt and customize our avatars. Six different avatars were created (see Figure \ref{teaser}) with random names including a female doctor (A) with medical clothes, a male doctor (B), a male (C) and a female (D) character with casual clothes, a female character associated with the Goth subculture (E), and a beaver avatar (F) with a male voice. Both doctors held the title "Dr." in their names, indicating their medical profession. The avatars were not created based on specific models, but rather inspired by the earlier related work on avatar appearance with regard to clothing, facial characteristics, hairstyle, ethnicity, age, gender, or a combination of these elements \cite{maria,56,markus}. The avatar voices were generated using an open-source text-to-speech platform\footnote{\url{https://ttsmp3.com/}} and the voice snippets were synchronized with the avatars using the \textit{SALSA Lipsynch} package within \textit{Unity} that provides automatic lip and emotion synchronization for digital objects and avatars.

\subsection{Questionnaires}

\textbf{PROM questionnaire:} The core questionnaire of the experiment consisted of different Patient-Reported Outcome items retrieved from PROMIS . PROM questionnaires are self-reported instruments used to measure patients' subjective experiences of their health status. These measures assess various aspects of health-related quality of life, such as symptoms, functioning, and well-being \cite{Nelson2015}. The core purpose of PROMs is to provide healthcare professionals with a patient-centered perspective on treatment outcomes \cite{Greenhalgh2018}. The purpose of using a PROM questionnaire was to simulate a health data measurement, which must be regarded as a particularly sensitive situation for data collection since it involves health data. 

%long form with items: The purpose of using a PROM questionnaire as the core questionnaire to be collected in the experiment was twofold: On the one hand, it forms the fictitious framework of a health data measurement, which must be regarded as a particularly sensitive situation for data collection, since it involves health data. On the other hand, the questionnaire we compiled contained all the items needed to create the global mental health score (GMHS). For this purpose, we included the items Global02 (In general, would you say your quality of life is…), Global04 (In general, how would you rate your mental health?), Global05 (In general, how would you rate your satisfaction with social activities and relationships?), Global10 (How often have you been bothered by emotional problems?) from the Global health item bank \cite{Hays2009}.

\textbf{Avatar rating questionnaires:} Participants were asked to rate their overall interaction and their willingness to share personal data, which is crucial for the PROM use case. Additionally, ratings on the avatar characteristics of warmth, competence, attractiveness, and human-likeness were collected. Our rating items were inspired by related work such as \cite{fiske1} with a certain degree of modification and adaptation to fit with our experiment. All the avatar rating items were scaled from 1 (\textit{Not at all}) to 5 (\textit{Absolutely}). Table \ref{predtable} lists all the questions used to rate avatar \textit{C}. At the end of the experiment, the participants were asked to fill out a small final questionnaire to pick their favorite avatar. We asked them to pick a final avatar they would prefer to interact with within a PROM application and add their feedback as comments.

\begin{figure}[htbp]
\centerline{\includegraphics[width=9cm, height=7cm]{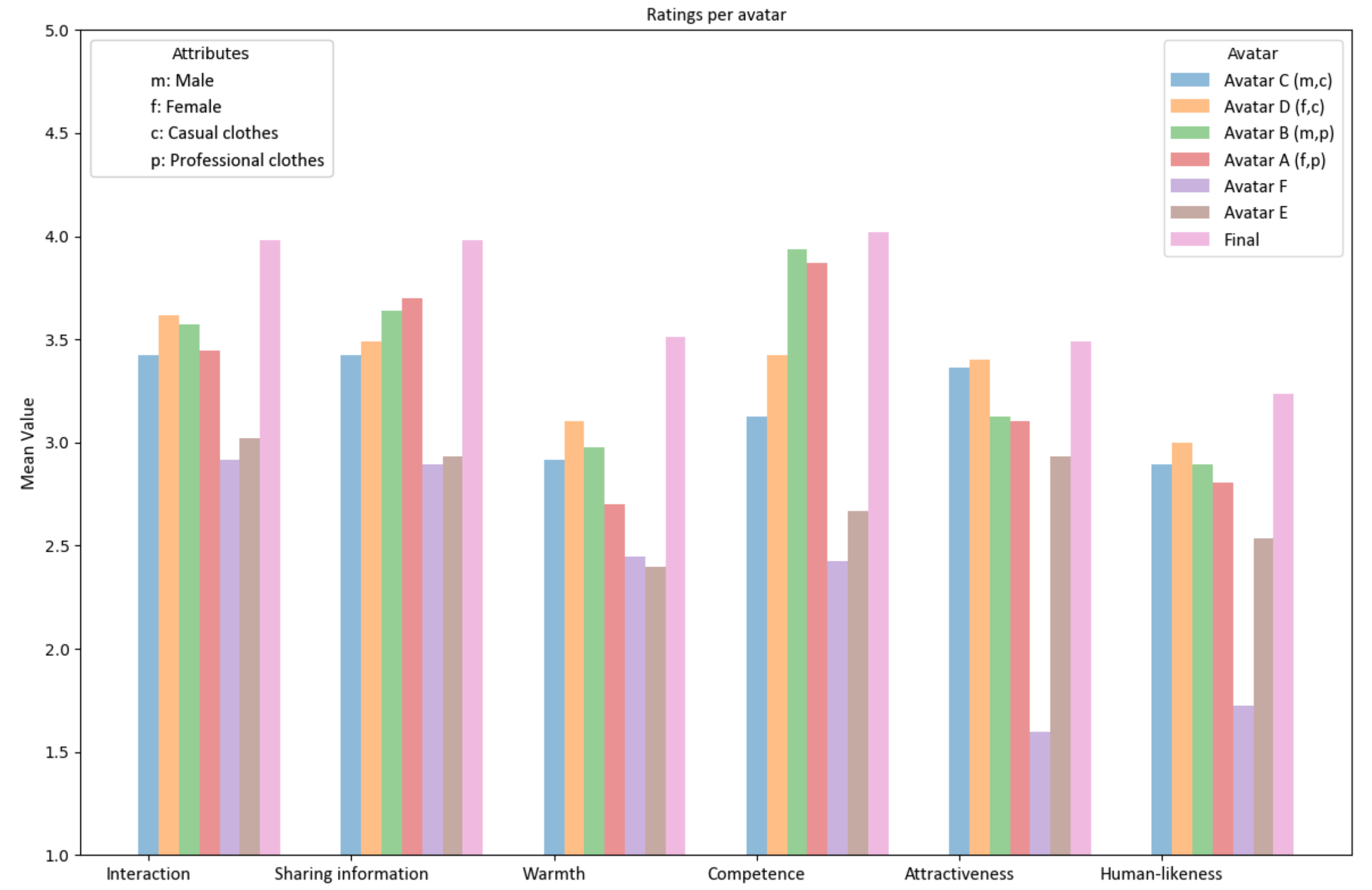}}
\centering
%\captionsetup{justification=centering}
\caption{Mean values over all recorded user inputs for each avatar rating plus mean values for the final avatar choices. The ratings belong to all six avatars with a specification of social role and gender.}
\label{Vis}
\end{figure}
\section{Results}
The results are divided into three sections, beginning with a report of results collected from the six virtual assistants. This is followed by the results from the chosen favorite virtual assistant and, finally, the results of the two-factor repeated measure ANOVA on gender (male/female) and clothing style (professional/casual), which included four out of the six avatars.  

\subsection{Descriptive Analysis of the Virtual Assistants' Rating Questionnaires}
While all avatars performed the same task in the same manner, the participants perceived them differently. Avatars \textit{A, B, C,} and \textit{D} – the anthropomorphic avatars in the norm range – received more positive feedback in general (Figure \ref{Vis}). Both doctor avatars, wearing medical-professional clothing, were considered the most competent \textit{(M=3.94, STD=0.99 and M=3.87, STD=0.90)}, while the simple male and female avatars were perceived as the most attractive \textit{(M=3.36, STD=1.01 and M=3.40, STD=1.01)}. The beaver received the lowest ratings in general, scoring particularly low on the attractiveness \textit{(M=1.60, STD 1.10)} and human-likeness \textit{(M=1.72, STD=0.88)} scales as one would expect. The biggest differences in the ratings can be observed for the attribute on the competence scale. Both avatars with a medical role received the highest scores, while the beaver and the subculture-associated avatar were rated lowest.

\subsection{Favorite Virtual Assistant choice}
As favorites, the female doctor agent was chosen with one-third most often (34\%, n=15), followed by the male doctor with almost a third (30\%, n=13). The last third comprises the two avatars in casual clothing - Avatar D (18\%, n=8), and Avatar (14\%, n=6). The beaver was the choice of two people (4\%), while the goth avatar was not chosen by anyone. Two participants did not specify any preference. The individual ratings of the avatars, when related to the favorite selected avatars, reveal more descriptive results about the importance of each scale when making a choice for this specific use case.

\begin{table}[tb]
    \caption{Items used in all avatar rating questionnaires (exemplary values are displayed for Avatar \textit{C}).\\
    \textbf{Note*:} All ratings were from a scale of 1 (\textbf{Not at all}) to 5 (\textbf{Absolutely}).}
    \label{tab:vis_papers}
  \scriptsize%
	\centering%
  \begin{tabu}{%
	l%
	*{7}{c}%
	*{2}{r}%
	}
        \hline
        Items & Mean & STD\\
        \hline
    I generally felt good interacting with David. & 3.43& 0.88 \\
    I felt comfortable sharing my personal information\\ with David.& 3.43 & 0.97\\
    How would you rate David's warmth?& 2.91& 1.14\\
    How would you rate David's competence?& 3.13& 1.01\\
    How would you rate David's attractiveness?& 3.36& 1.01\\
    Did you feel a human-like connection while listening\\ to David?& 2.89& 1.18\\
        \hline
    \end{tabu}
    \label{predtable}
    
    %\end{center}
\end{table}
\begin{figure}[htbp]
\includegraphics[width=12cm, height=12cm]{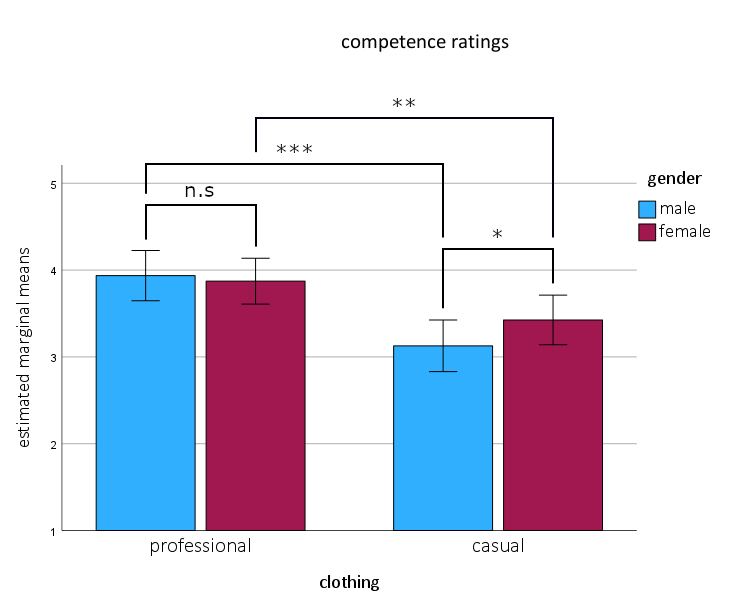}
\centering
%\captionsetup{belowskip=0pt}
%\captionsetup{justification=centering}
\caption{Mean comparisons of avatars' competence ratings based on clothing style (professional vs. casual) and gender (male vs. female). The bars represent estimated marginal means with error bars indicating standard errors. Pairwise comparisons between conditions are marked with significance levels: (*** p\textless 0.001 ** p\textless 0.01 * p\textless 0.05 n.s. not significant). The results indicate a significant difference in competence ratings between professional and casual clothing, as well as between certain gender and clothing combinations.}
\label{comptence}
\end{figure}
%\vspace{-10pt}
Both, the ratings of the overall interaction \textit{(M=3.98, STD=0.82)} and the willingness to share information \textit{(M=3.98, STD=0.77)} were rated high for the favorite avatar. Among the avatar characteristics, the competence scale turned out to be the most relevant for the PROM use case \textit{(M=4.02, STD=0.97)}, while warmth \textit{(M= 3.51, STD=1.06)} and attractiveness \textit{(M=3.49, STD=0.98)} are in the middle range, and human-likeness \textit{(M=3.23, STD=1.20)} was rated lowest. The distribution of ratings for the avatar chosen as favorite is depicted in Figure \ref{Vis}.

Regarding user characteristics and characteristics of the chosen avatar, the opposite gender was slightly preferred, with around two-thirds of each gender picking an avatar with the opposite gender as theirs. This trend, however, was not statistically significant (p=0.097). A slight trend also revealed that men tended to prefer avatars with a professional appearance (80\%:20\%), whereas women equally chose between an avatar in a professional or non-professional role. This trend, although not statistically significant, presents an interesting point of discussion (p=0.076).

\subsection{Repeated Measure ANOVA}
This section reports the findings from the repeated measures ANOVA. Here, no significant differences emerged relating to gender. However, a significant difference was found in terms of clothing and an interaction effect was observed between gender and clothing on the competence scale. Please note that the ANOVA focused on only four avatars and showed significant effects on competence-rating.
A two-factor analysis of variance with repeated measures revealed that, multivariately, the main effect of clothing style (F(6, 41) = 8.481, p = 0.001, \( \eta p^2 \) = 0.554) was significantly associated with the ratings of the virtual agents, while the gender of the avatars did not show significance (F(6, 41) = 0.598, p = 0.730, \( \eta p^2 \)= 0.080). The interaction effect between clothing style and gender also demonstrated a significant association (F(6, 41) = 2.439, p = 0.042, \( \eta p^2 \) = 0.263). Univariate analyses revealed the following significant relationships: competence ratings (F = 25.628, p = 0.001, \( \eta p^2 \) = 0.358) and attractiveness ratings (F = 5.380, p = 0.025, \( \eta p^2 \) = 0.105) for the clothing factor, and competence ratings (F = 5.790, p = 0.020, \( \eta p^2 \) = 0.112) for the interaction of clothing and gender. The warmth rating also approached significance for the interaction effect of clothing and gender (F = 3.343, p = 0.074, \( \eta p^2 \) = 0.068). 

All other relationships were not significant. Figure \ref{comptence} depicts the estimated marginal means of the two factors with pairwise comparisons, showing significant differences between casually dressed male and female avatars (p = 0.025), professionally and casually dressed females (p = 0.002) and the highest difference between professionally and casually dressed males (p = 0.001), while the mean difference between professionally dressed males and females is not significant (p = 0.537).
In contrast, the pairwise comparisons for warmth reveal significant differences between professionally and casually dressed women (p = 0.045).

The significant interaction between gender and clothing shows that the difference in competence between the two clothing options is different for men and women. Pairwise comparisons show that the difference in competence between the two clothing options is greater for men than for women. This indicates that clothing has a stronger influence on perceived competence for men than for women. The opposite is true for the ratings of warmth - clothing has a stronger influence on perceived warmth for women than for men.

\subsection{Analysis of the open questions}
In total 19 participants left 49 comments – most of them on the overall study and regarding factors for the choice of preference. The analysis revealed more variation than in the statistical analysis, which is in line with Human-Computer Interaction (HCI) literature: Concrete factors or characteristics that are liked by some are disliked by others. For example, while most people disliked artificial voices, one person found a 'robot-like' voice competent and fitting for the purpose, while others experienced it as a fit for an artificial agent. The feedback questions also show that great attention is paid to gaze behavior and overall appearance. Both aspects are evaluated regarding the criterion of liveliness, which ranges between a natural and warm appearance with positive connotations and a machine-like and artificial appearance perceived as cold and probably uncanny. Some comments revealed that experience with other voice assistance systems or avatars has become an important reference in the evaluation of interactions with such systems. These were critical comments that attested to a lack of quality in the evaluated system compared to the systems they were already familiar with. 

\section{Discussion}
The significant variations in ratings across individual scales highlight the diversity in people's perceptions of the different avatars. This can also be seen in the choice of the preferred avatar since no clear preference can be observed except that just two participants chose avatar \textit{F}, and no one chose avatar \textit{E}. The latter is attributed to the fact that the association with the "goth" subculture was perceived as not fitting in the context of medical applications. Interestingly, this does not seem fully clear for an animal character. We attributethe high scores for the avatars \textit{A, B, C}, and \textit{D} on the warmth and humanoid scale to the fact that the external appearance of these four VAs remains within the ‘norm' range people are familiar with and would prefer in a professional context. Accordingly, the fact that the beaver avatar scored the lowest on the attractiveness and human-likeliness scales can be attributed to the fact that these scales ask for decidedly human characteristics that do not fit other than human characters, nonetheless, we have used the same rating items for the beaver to keep the coherence in our collected data.
Regarding the ratings of the favorite avatar, it can be noted that the aspects of competence, sharing information, and overall interaction with an avatar are more important in choosing a preferred VA than warmth, attractiveness, and human-likeness. This makes sense, as these are especially human characteristics that only play a subordinate role for an avatar, especially in the actual medical setting. In particular, the human-likeliness score stands out, although this must also be attributed to the fact that none of the VAs achieved a high human-likeliness rating. This also shows that human-likeliness is not a highly relevant factor. Even though avatars \textit{E} and \textit{F} received similar (low) ratings – especially compared to the others –, it is noticeable that the goth avatar was not chosen once as a favorite. We acknowledge that more complex factors than the ones assessed in this study may have influenced the choice of a favorite avatar.

The results of the repeated measure ANOVA reveal significant differences in clothing style but not in gender in the competence ratings. Interestingly, the interaction effect of clothing and gender is also significant, which means that male vs. female professionally dressed avatars are rated differently from male vs. female casually dressed agents. Looking into the pairwise comparisons, the biggest competence difference is between the male professional and the male casual avatar. The results show that male professional avatars are perceived more positively here. We attribute this difference to gender-stereotypical patterns of perception and evaluation. They likely benefit from the combination of their characteristics, which seems to correspond to the socially expected context. Unsurprisingly, there is an analogous correlation between the female gender and the attribution of warmth and attractiveness in that the avatar of the professional woman was rated lower in these characteristics. We see the same stereotypical correlation here but extended by one dimension: the role of the professional woman is perceived as less pleasant (warm) because it does not seem to correspond exactly to the social image of women and, therefore, cannot be associated with the same positive experiences and evaluations.

The preference for a female avatar in the role of the medical expert confirms the results of other empirical studies on this topic \cite{TerStal2020}. The attributes warm and cold are among the most important core attributes of the construction of gender, with warmth usually associated with the female gender and cold usually associated with the male gender \cite{Hochschild1995}. Taking this into account, we can assume that the strong preference of men for choosing a female avatar can be attributed to the warmth that is usually associated with this gender. Thus, a slight correlation between the need for warmth and avatar preferences could indeed be noted, although we would like to point out here that gender attributes must be considered as socially constructed \cite{Butler2004} and their reproduction must be viewed critically \cite{Bardzell2010}. However, while these and other studies \cite{hallqvist2019digital,ter2020embodied} assign a slightly higher influence to gender than to the professional role, we find a more substantial influence on the medical role. This is in line with the competence rating being the most significant factor on our measurement scales.

\section{Conclusion}

Virtual avatars are turning into a significantly popular component of healthcare applications and can directly influence the overall QoE. Therefore, analyzing the underlying factors that lead to particular avatar preferences are of paramount importance in the realm of user-centered healthcare applications. In this study, we showed that the perception of virtual avatars can strongly vary among participants and no strict guidelines could be defined for designing virtual agents for medical applications. We have highlighted the importance of certain characteristics of the avatars, such as competence and information sharing, over warmth, attractiveness, and human-likeness, in influencing avatar preferences. This study sheds light on how competence ratings are affected by biases in perception, which might arise from socially ingrained gender stereotypes and clothing styles. Notably, male and female avatars dressed professionally are perceived differently due to distinct social expectations; the professional male avatar is assigned higher competence ratings, while the female avatar attains lower evaluations of attractiveness and warmth. It thus becomes clear that culturally indoctrinated gender-stereotypical biases persist, affecting perceptions even within the context of virtual avatars, underscoring the necessity for increased consciousness and mitigation of such biases.
\bibliographystyle{splncs04}
\bibliography{samplepaper}
%Author, F.: Article title. Journal \textbf{2}(5), 99--110 (2016)

%Author, F., Author, S.: Title of a proceedings paper. In: Editor,
%F., Editor, S. (eds.) CONFERENCE 2016, LNCS, vol. 9999, pp. 1--13.
%Springer, Heidelberg (2016). \doi{10.10007/1234567890}

%Author, F., Author, S., Author, T.: Book title. 2nd edn. Publisher,

%Author, A.-B.: Contribution title. In: 9th International Proceedings
%on Proceedings, pp. 1--2. Publisher, Location (2010)

%LNCS Homepage, \url{http://www.springer.com/lncs}, last accessed 2023/10/25
%\end{thebibliography}
\end{document}